# Unveiling the impact of the bias-dependent charge neutrality point on graphene-based multi-transistor applications

Francisco Pasadas[1,3], Alberto Medina-Rull[2], Pedro Carlos Feijoo[1], Anibal Pacheco-Sanchez[1], Enrique G. Marin[2], Francisco G. Ruiz[2], Noel Rodriguez[2], Andrés Godoy[2] and David Jiménez[1]

[1] Departament d'Enginyeria Electrònica, Escola d'Enginyeria, Universitat Autònoma de Barcelona, Bellaterra, Spain
[2] Departamento de Electrónica y Tecnología de Computadores, PEARL Laboratory, Facultad de Ciencias, Universidad de Granada, Granada, Spain
[3] Author to whom any correspondence should be addressed

E-mail: francisco.pasadas@uab.es

## Abstract

The Dirac voltage of a graphene field-effect transistor (GFET) stands for the gate bias that sets the charge neutrality condition in the channel, thus resulting in a minimum conductivity. Controlling its dependence on the terminal biases is crucial for the design and optimization of radio-frequency applications based on multiple GFETs. However, the previous analysis of such dependence carried out for single devices is uncomplete and if not properly understood could result in circuit designs with poor performance. The control of the Dirac point shift (DPS) is particularly important for the deployment of graphene-based differential circuit topologies where keeping a strict symmetry between the electrically balanced branches is essential for exploiting the advantages of such topologies. This note sheds light on the impact of terminal biases on the DPS in a real device and sets a rigorous methodology to control it so to eventually optimize and exploit the performance of radio-frequency applications based on GFETs.

## 1. Introduction

Research in graphene-based electronics has found in radio-frequency (RF) applications an open and fertile land where to exploit the carbon allotrope superlative transport properties, i.e. ultra-high carrier mobility ($2\times10^5$ cm$^2$/Vs) and a high saturation velocity (~$5.5\times10^7$ cm/s) [1–7]. The field-effect transistor (FET) is an important component of any RF circuit embedded in the modern information and communication systems, and thus, improvements of its performance have a critical impact on this field [7–9]. Indeed, very auspicious RF graphene field-effect transistors (GFETs) have already been demonstrated [10–12]. An intrinsic (after a de-embedding procedure) cutoff frequency of 427 GHz measured in a graphene-based FET (GFET) of 67-nm channel length is the highest reported so far [13]. Similarly, an intrinsic maximum oscillation frequency of 200 GHz has been achieved in a GFET of 60-nm length [14]. Although further improvement of the RF device performance is expected [15], the already proven GFET figures of merit (FoMs) allow the design of applications operating at millimeter wave (mmWave) frequencies. Those encompass applications where the GFET leverages either its remarkable high transconductance, e.g., low noise amplifiers (LNAs) [16]; or the graphene ambipolarity such as resistive mixers [17–21]; frequency multipliers [21,22]; and power detectors [23], all of them demonstrated to properly operate in the gigahertz band.

To push graphene electronics towards higher technology readiness levels (TRLs), and shaping GFETs as the core of novel RF applications, it is first necessary to deeply comprehend those factors that more critically impact the circuit performance [10–12]. Among them, the Dirac point shift (DPS) is crucial in defining the circuit performance. Specifically, applications based on graphene ambipolarity need a tight control of the DPS. To this purpose, the inclusion of a back-gate in the GFET architecture has been the standard strategy followed so far. Numerous examples have been reported such as the polarity-controllable graphene inverter, the voltage controlled resistor [24,25], or the graphene-based frequency tripler [26] and quadrupler [27] successfully implemented thanks to a carefully adjusted voltage separation of two cascaded GFETs. However, a more advanced graphene technology might require to abandon the four-terminal architecture because of the higher complexity inherent to the fabrication process and, more importantly, the appearance of parasitics due to the back gate, seriously compromising the high-frequency performance [28]. Hence, a thorough understanding of the bias-dependent DPS is required for circuit design, an example of which will be presented to obtain frequency multiplication with single-gated GFETs.

On the other hand, the vast majority of experimental GFET-based circuits reported so far are founded on single-transistor stages. However, the better performing circuits to come will rely on multiple-transistor stages, where the direct observations extracted from single devices become inappropriate. This work shows that a careful consideration of the DPS bias dependence is necessary to reach the optimal performance of multi-transistor circuits. Specifically, the DPS experienced by each GFET in a differential topology can ruin the required symmetry, spoiling their performance. This deleterious effect has been reported, e.g., in [29] where a dissymmetry under large drain-to-source biases appeared between the two GFETs comprising a balun. In addition, a different DPS rate from the usually expected, i.e. half of the externally applied drain-to-source bias, has been observed in [30] for two GFETs connected in a complementary-like inverter topology. It is in this forthcoming technological arena where the DPS can hamper the performance of multi-GFET circuits; an issue that we intend to describe and rationalize in this note, which is structured as follows: Section 2 presents the theory behind the DPS, by analyzing the electrostatics, and a comparison of the predictions against dc measurements of a GFET reported elsewhere [31]; Section 3 describes how to control the impact of the DPS in a multi-transistor circuit while Section 4 shows the development of new functionalities making use of a deeper understanding of the DPS phenomenon. Finally, the conclusions are drawn in Section 5.

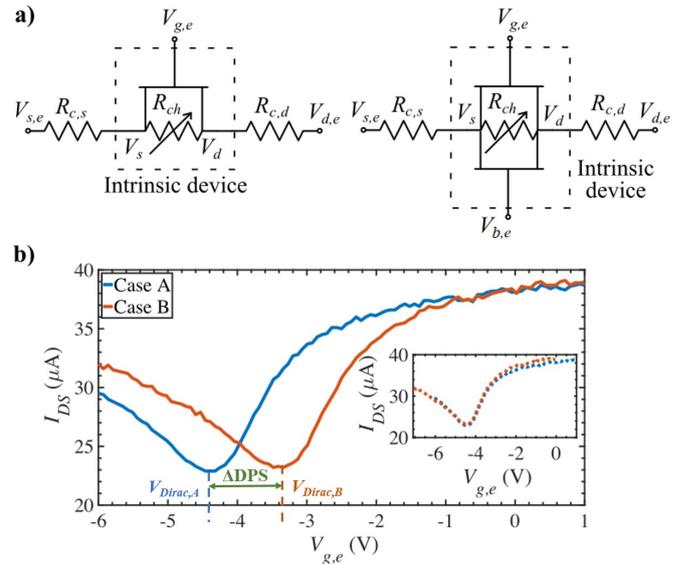

**Figure 1.** a) Schematics of the three- (left) and four-terminal (right) GFET resistive networks. The electrostatic modulation of the bias-dependent channel resistance ($R_{ch}$) is achieved via the gate terminal(s) ($V_{g,e}$ and/or $V_{b,e}$). Due to the non-negligible metal-graphene contact resistances ($R_{c,s}$ and $R_{c,d}$), a substantial potential drop is produced so that the intrinsic voltages ($V_s$ and $V_d$), not accessible in practice, are quite different from the external ones ($V_{s,e}$ and $V_{d,e}$). b) Measurements of the dc transfer characteristics of the GFET reported in [31] at two different operating bias point: (Case A) $V_{d,e} = 0.1$V and $V_{s,e} = 0$V; and (Case B) $V_{d,e} = 1.1$V and $V_{s,e} = 1$V, showing a different DPS despite considering the same $V_{ds,e}$ in both cases ($\Delta$DPS=$V_{Dirac,A}$−$V_{Dirac,B}$). (inset) Measured transfer characteristics after countering the DPS according to the model proposed in eqs. (2) and (3).

## 2. Dirac Point Shift (DPS): theory *vs*. experiment

The Dirac voltage ($V_{Dirac}$) of a GFET is defined as the gate bias that sets the charge neutrality condition in the graphene channel, namely, the gate bias that results in a minimum conductivity, and therefore, it can be readily located at the vertex of the V-shaped transfer characteristics (TCs) of the GFET. For a well-behaved long-channel GFET, we can estimate $V_{Dirac}$ by considering same electron/hole mobility as well as symmetrical charge distribution, i.e., same electron/hole concentration at the source and drain edges. These concentrations can be obtained upon application of Gauss' law along a vertical cut of the GFET structure, i.e., assuming the gradual channel approximation, and after considering the quasi-Fermi levels equal to $V_s$ and $V_d$ at the source and drain terminals, respectively [32]. For ultra-short-channel devices, the 2D Poisson equation should be considered instead [33]. Aiming to get a compact equation that could be easily handled by circuit designers, only the long-channel case has been considered in this work.

Using the abovementioned methodology, the $V_{Dirac}$-terminal-bias dependence of a four-terminal GFET can be written as:

$$V_{Dirac} = V_{g0} + \left(\frac{C_t + C_b}{C_t}\right)\frac{(V_d + V_s)}{2} - \frac{C_b}{C_t}(V_b - V_{b0}), \qquad (1)$$

where $C_t$ and $C_b$ are the top and bottom oxide geometrical capacitances per unit area, respectively, and $V_g$-$V_{g0}$ and $V_b$-$V_{b0}$ are the top and bottom gate voltage overdrive, respectively. These quantities embrace work-function differences between the gates and the graphene channel and the possible presence of additional charges due to impurities or doping. The dependence of $V_{Dirac}$ in eq. (1) on $V_d$ and $V_s$ is a unique feature in devices with ambipolar transport [34], and has also been observed in carbon nanotube devices [35,36]. Indeed, the term $(C_b/C_t)(V_b-V_{b0})$ in eq. (1) has been usually used to experimentally obtain the ratio $(C_b/C_t)$ by sweeping $V_b$ while keeping the rest of terminal biases constant [37–39].

If one of the gate capacitances per unit area, e.g. the top-gate, is much larger than the other, i.e. $C_t \gg C_b$, then the four-terminal GFET can be considered as a three-terminal device and eq. (1) simplifies into:

$$V_{Dirac} = V_{g0} + \frac{(V_d + V_s)}{2}. \qquad (2)$$

Most of experimental GFETs found in the literature are either three-terminal devices or four-terminal ones where the condition $C_t \gg C_b$ is met, and therefore, eq. (2) commonly applies. Remarkably, the bias dependence of $V_{Dirac}$ reported in the literature [29,30,40–42] is in many cases successfully connected to the external drain and source biases, $V_{d,e}$ and $V_{s,e}$, respectively, specifically to $(V_{d,e} - V_{s,e})/2 = V_{ds,e}/2$, instead of $(V_d + V_s)/2$ predicted in eq. (2) (figure 1a schematically depicts the difference between external and intrinsic voltages). We have found out the reason of such a coincidence and carried out measurements to demonstrate that eq. (2) should be considered instead.

Figure 1a shows the equivalent resistive networks of both three- and four-terminal GFETs with the intrinsic graphene channel resistance ($R_{ch}$) and the extrinsic drain ($R_{c,d}$) and source ($R_{c,s}$) resistances. These include the drain and source metal-graphene contact resistances, respectively, which are currently crucial and undesirable elements impacting the RF performance of GFETs [43–52]. For the sake of simplicity, bias-independent contact resistances have been considered in the device modeling approach of this work. The bias dependence of physical effects at metal-graphene interfaces affecting the carrier transport [46,51] can be embraced by the channel charge description as incorporated by the device model used here [53]. Assuming that the drain/source extrinsic resistances are similar $R_{ext} = R_{c,d} \approx R_{c,s}$, then the intrinsic drain and source voltages ($V_d$ and $V_s$, respectively) can be determined as:

$$\begin{aligned} V_d &= V_{d,e}\frac{1+x}{2+x} + V_{s,e}\frac{1}{2+x}, \\ V_s &= V_{d,e}\frac{1}{2+x} + V_{s,e}\frac{1+x}{2+x}, \end{aligned} \qquad (3)$$

where $x = R_{ch}/R_{ext}$. Interestingly, from eq. (3) we observe, adding up both expressions, that the equality $(V_d + V_s) = V_{d,e} + V_{s,e}$ holds and as $V_{s,e}$ is usually shorted to ground, then $(V_d + V_s) = V_{ds,e}$. However, this is not a general case, figure 1b shows a proper counterexample: the TCs of a 14-μm length GFET reported in [31] in two scenarios: (Case A) $V_{d,e} = 0.1$V and $V_{s,e} = 0$V; and (Case B) $V_{d,e} = 1.1$V and $V_{s,e} = 1$V; thus keeping in both cases same $V_{ds,e} = 0.1$V, but in case B breaking the condition of $V_{s,e}$ grounded. As it can be seen, in spite of what has been so far widely reported [29,30,40–42], the DPS indeed depends on $(V_d + V_s)$, as the same $V_{ds,e}$ cannot explain these two different DPSs. The inset in figure 1b shows the measured TCs after countering the DPS according to eqs. (2) and (3), supporting the exposed theory and demonstrating a perfect cancellation of the shift.

More importantly, because of the presence of non-negligible extrinsic resistances in GFETs, the potential $V_s$ cannot be shorted to reference in practice. Indeed, recent studies on contact resistance of GFETs [54,55] show that the ratio $x = R_{ch} / R_{ext}$ is lower than 1 for the vast majority of experimental devices, meaning that most of the external $V_{ds,e}$ drops at the extrinsic resistances. Thus any change in either $V_{g,e}$ or $V_{d,e}$ would cause a change in $V_s$, revealing that the DPS needs to be taken into account in the circuit design.

## 3. A case of practice: impact of DPS on an RF balun

Differential topologies could become attractive candidates for the development of high-performance RF graphene applications as they provide less even-order distortion, immunity to common-mode noise couplings and crosstalk through the substrate and supply rails [56,57]. However, such topologies are essentially based on the symmetry of the electrically balanced branches and ultimately relied on the symmetry of the transistor outcome. For the specific case of differential pairs, the symmetry has been demonstrated to be essential and analyzed from a mathematical perspective, reveling its impact on the performance of this amplifier topology [58,59].

The results shown in Section 2 lead us to investigate the unavoidable DPS impact on the RF performance of GFET-based applications with symmetry requirements. A paradigmatic example arises when designing a balun. This circuit, shown in figure 2a, is central for the development of graphene-based differential circuit topologies, exploiting the unique symmetrical and ambipolar behavior of graphene. In most cases, the incoming signals originated in antennas are single-ended and to achieve the differential topology

TABLE I
GRAPHENE TECHNOLOGY CONSIDERED FOR THE CIRCUIT DESIGN

| $L$ | 1 μm | $W$ | 1 μm | $R_g L$ | 5 Ωμm |
|---|---|---|---|---|---|
| $\mu$ | 0.2 m$^2$/Vs | $V_{g0}, V_{b0}$ | 0 V | $R_{d1}, R_{d2}$ | 4 kΩ |
| $C_t$ | 8 fF/μm$^2$ | $C_b$ | $C_t$/231 | $R_s, R_d$ | 200 Ωμm |

*$\mu$ represents the effective carrier mobility and $R_g$ the gate resistance.

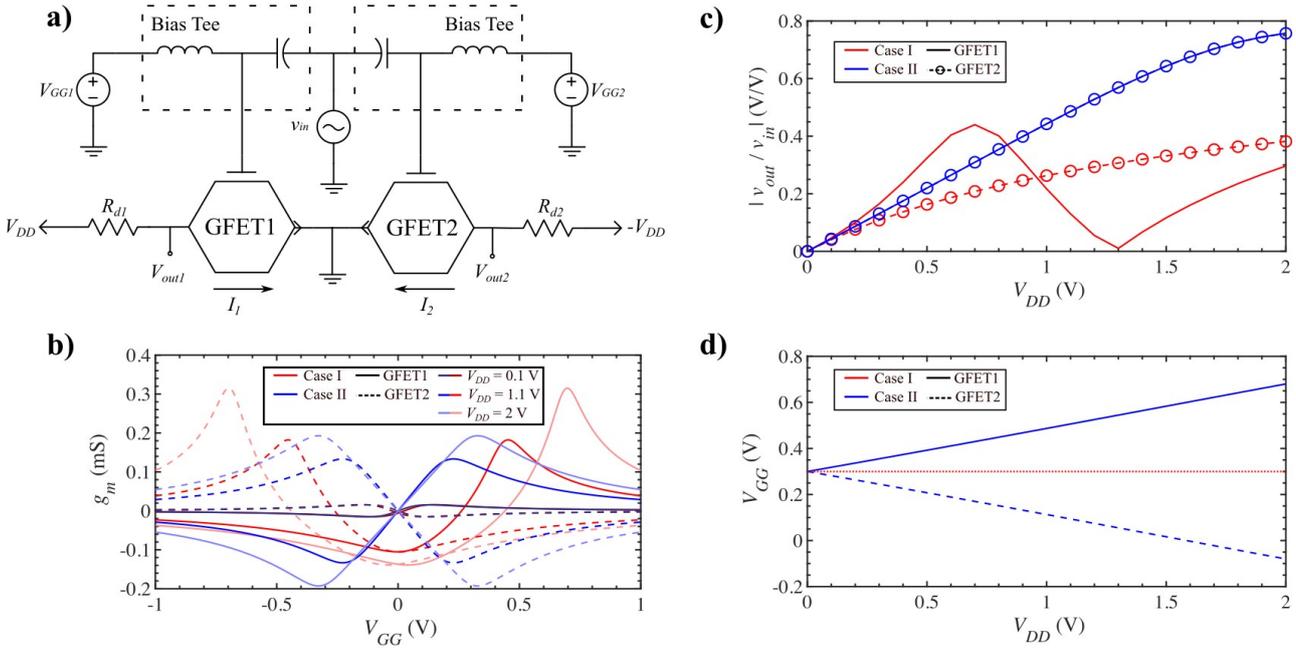

**Figure 2.** a) Schematics of a graphene-based balun [29,57]. In order to automatically counter the DPS, we propose to use different dc paths for feeding the gate of each device. b) Transconductance ($g_m = \partial I/\partial V_{GG}$) of each GFET showing the DPS effect (Case I) at constant $V_{GG}$ and (Case II) countering the DPS effect by properly changing $V_{GG1}$ and $V_{GG2}$ for different $V_{DD}$. c) Voltage gain ($|v_{out}/v_{in}| = g_m R_d$) and d) corresponding gate bias applied to each GFET for both Case I and II.

concept, a balun circuit is required to transform the single-ended signal ($v_{in}$, unbalanced) into the differential signals ($V_{out1}$ and $V_{out2}$, balanced, i.e., with a π-rad phase difference). To take full advantage of differential topologies, the power of both balanced signals should be similar. However, for a graphene-based circuit, the inherent DPS will break the symmetry between both transistors, making mandatory to carefully counter-balance it by continuously applying effective gate biases to each GFET as described by eq. (2). Notice that the bias tees have been used in the schematics (figure 2a) for pointing out explicitly different dc effective gate bias for each device. However, this would not spoil the dc power consumption because the gate terminals behave ideally as open circuits. Considering a specific operating bias point, both GFETs could, in this context, be fed by a single power supply through the use of a voltage divider.

The design and analysis of an RF balun based on graphene, requires a physics-based description of the electrical behavior of the GFET. To this purpose, we have embedded into Keysight© ADS, a large-signal model implemented in Verilog-A by some of the authors [53] which has been demonstrated to show good agreement with measurements [60]. The main features of the graphene technology considered are summarized in Table I.

Figure 2b depicts the transconductance ($g_m = \partial I/\partial V_{GG}$) of each GFET (labeled 1 and 2) without countering the DPS effect (Case I) at constant $V_{GG}=V_{GG1}=V_{GG2}$ and countering the DPS effect (Case II). The control of the DPS can be realized by tuning $V_{GG1}$ and $V_{GG2}$. In Case I (figure 2b), when $V_{DD}$ is low, the DPS does not substantially break the symmetry and, thus, both transconductances have similar modulus. However, if $V_{DD}$ is large enough, the symmetry between both devices is broken and, for a given $V_{GG}$, $|g_m|$ for each GFET is considerably different, resulting in unbalanced output signals (not shown here) and leading to a useless circuit. As a result, the DPS in Case I limits the use of this circuit only for low drain biases (see figure 2c), resulting in strongly attenuated output signals, as exemplified by the -20dB gain values obtained by the graphene-based balun reported in [57]. On the other hand, when the DPS is counter-balanced (Case II), the balun proper operation is restored given that the recovered symmetry leads to similar $|g_m|$ for both GFETs (cf. figure 2b). In addition, increasing $V_{DD}$ in figure 2c leads to the desired voltage gain for both balanced signals through the automatic application of different gate biases (by countering the DPS), e. g., linearly with $V_{DD}$, but keeping opposite slopes as shown in figure 2d. This way, given that the DPS is under control, the supply bias ($V_{DD}$) offers an additional degree of freedom to tune the desired voltage gain, although at the cost of a higher dc power consumption.

## 4. Exploiting DPS: frequency multiplication based on single-gated multi-GFET stage

In this Section, the previous analysis of the DPS phenomenon allows us to propose a solution to achieve frequency multiplication with single-gated GFETs.

The working principle of the frequency tripler [26] and quadrupler [27] is based on a W-shaped TC observed in two cascaded GFETs that can be achieved by controlling individually the applied electrical field dependence of $V_{Dirac}$ of each device (e.g., see figure 3d). Up to now, the separated electrostatic control of $V_{Dirac}$ has been demonstrated by the use of dual-gated devices (i.e., by adjusting $V_b$ in eq. (1)) [26,27]. However, according to eq. (2), it is also possible to modify the $V_{Dirac}$ of a single-gated device by tunning its source and drain voltages, avoiding the introduction of a fourth terminal. Therefore, the search of W-shaped TCs in single-gated GFETs is of upmost interest, and of partial success in [61] where W-shaped responses were observed, although in an uncontrolled way as illustrated by the fact that devices with the same structure and dimensions presented different behaviors [61]. Here, leveraging the understanding of the DPS behavior aforementioned, we have implemented an original and simplified way to achieve W-shaped TCs in single-gated devices to exploit frequency multiplication based on graphene ambipolarity.

To do so, a lumped resistor ($R_X$) is included between the two cascaded GFETs as shown in figure 3a in order to produce a controlled separation of the DPS of each GFET (proportional to the actual voltage drop at $R_X$) giving rise to a W-shaped TC that can be exploited for frequency multiplication.

Figure 3a depicts the proposed single-gated GFET-based frequency multiplier, whose TCs are represented in figure 3b-c. Specifically, the W-shaped TC tunability with both the value of the resistor $R_X$ for a given supply bias and the value of the supply bias ($V_{DD}$) for a given $R_X$ are depicted, respectively. These results indicate that the voltage drop originated at $R_X$ causes an increasing separation between the charge neutrality points of both GFETs. Figure 3d shows a sketch of the working principle for a frequency tripler and quadrupler based on the W-shaped TC at $V_{DD} = 2V$ with $R_X = 1k\Omega$ (green curve in figure 3c). Considering an input frequency of $f_{in} = 1$MHz, the analysis of the output power spectrum (not shown here) indicates that the resulting RF relative power for the frequency tripler is 61.9% at the triple fundamental frequency ($3f_{in}$), while the quadrupler shows a relative power of 38.6% at the quadruple fundamental frequency ($4f_{in}$). Insets in figure 3d depicts a comparison between the predicted output signals and pure sinusoids of corresponding frequencies showing an evident distortion. The performance of GFET-based frequency multipliers can be improved by additional matching and stability networks; however, this is out of the scope of this study intended to show the impact of the DPS control at a device level on multi-transistor configurations. Therefore, this design gives insight about how the accurate control of the bias-dependence of $V_{Dirac}$ can be exploited for the development of sophisticated RF applications.

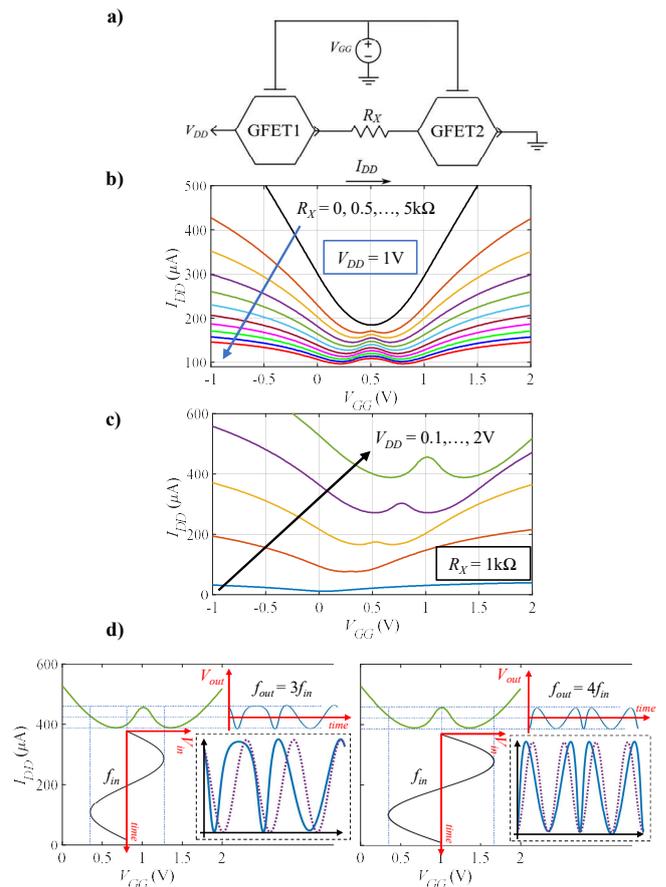

Figure 3. a) Schematics proposed to build graphene-based frequency multipliers taking advantage of the DPS phenomenon understanding. Each GFET is described in Table I. We propose the inclusion of a resistor, $R_X$, between two cascaded GFETs so to originate W-shaped transfer characteristics in single-gated devices which can be leveraged to frequency multiplication. Transfer characteristics for different b) values of $R_X$ (at $V_{DD}$ = 1V); and c) supply biases, $V_{DD}$ (considering $R_X$ = 1k$\Omega$). d) Working principle for the GFET-based frequency tripler (left) and quadrupler (right) based on the TC at $V_{DD}$ = 2V and $R_X$ = 1k$\Omega$. Inset: blue solid lines show the output signals of the frequency tripler (left) and quadrupler (right) compared against pure sinusoids of corresponding frequencies depicted by purple dashed lines.

## 5. Conclusion

This work demonstrates that the DPS in GFETs is proportional to the intrinsic voltages ($V_d + V_s$), not accessible in practice, which can be considerably different from the externally applied biases due to the non-negligible metal-graphene contact resistances. It has been shown that the validity of the usual approximation that considers the DPS proportional to $V_{ds,e}$, is limited and should be avoided.

Thus, a simple analytic equation for long-channel GFETs has been provided in order to evaluate the bias-dependence of DPS that could be compensated by the automatic application of an effective gate bias. As a proof of concept, the operation of a graphene-based balun that harnesses the unique ambipolar characteristic of graphene has been assured by controlling the DPS showing an improved performance. In addition, a frequency tripler/quadrupler based on W-shaped transfer characteristic of two cascaded single-gated GFETs is proposed, evidencing that monitoring the bias-dependence of the DPS is pivotal for the design of RF applications. In summary, this work provides critical insights on DPS phenomenon paving the way towards the development of RF multi-transistor circuits based on graphene technology.

## Acknowledgements

The authors would like to thank W. Kim, J. Riikonen and H. Lipsanen from Aalto University (Finland) for providing the dc measurements of the transfer characteristics shown in figure 1b. This work is supported in part by the Spanish Government under the projects TEC2015-67462-C2-1-R (MINECO), TEC2017-89955-P and RTI2018-097876-B-C21 (MCIU/AEI/FEDER, UE); the FEDER/Junta de Andalucía Consejería de Economía y Conocimiento under project B-RNM-375-UGR18; European Union's Horizon 2020 Research and Innovation Programme under Grant Agreement No. GrapheneCore3 881603. E.G. Marin gratefully acknowledges Juan de la Cierva Incorporación IJCI-2017-32297 (MINECO/AEI). F. Pasadas and D. Jiménez acknowledge the support from the ERDF allocated to the Programa Operatiu FEDER de Catalunya 2014-2020, with the support of the Secretaria d'Universitats i Recerca of the Departament d'Empresa i Coneixement of the Generalitat de Catalunya for emerging technology clusters to carry out valorization and transfer of research results. Reference of the GraphCAT project: 001-P-001702.

## Conflicts of interest or competing interests

The authors declare that they have no known competing financial interests or personal relationships that could have appeared to influence the work reported in this paper.

## Data availability

The data that support the findings of this study are available upon request from the authors.

## ORCID iDs

Francisco Pasadas https://orcid.org/0000-0003-3992-9864
Alberto Medina-Rull https://orcid.org/0000-0003-4691-0328
Pedro Carlos Feijoo https://orcid.org/0000-0002-7653-4573
Anibal Pacheco-Sanchez https://orcid.org/0000-0002-0897-0605
Enrique G. Marin https://orcid.org/0000-0002-0302-3764
Francisco G. Ruiz https://orcid.org/0000-0003-4659-2454
Noel Rodriguez https://orcid.org/0000-0002-6032-6921
Andrés Godoy https://orcid.org/0000-0002-3014-8765
David Jiménez https://orcid.org/0000-0002-8148-198X

## References

[1] Avouris P 2010 Graphene: electronic and photonic properties and devices *Nano Lett.* **10** 4285–94

[2] Castro Neto A H, Peres N M R, Novoselov K S and Geim A K 2009 The electronic properties of graphene *Rev. Mod. Phys.* **81** 109–62

[3] Liao L and Duan X 2012 Graphene for radio frequency electronics *Mater. Today* **15** 328–38

[4] Palacios T T, Hsu A and Wang H 2010 Applications of graphene devices in RF communications *IEEE Commun. Mag.* **48** 122–8

[5] Schwierz F and Liou J J 2003 *Modern Microwave Transistors: Theory, Design, and Performance* (Wiley)

[6] Taur, Yuan and Ning T H 2005 *Fundamentals of Modern VLSI Devices* (New York: Cambridge Univ Press)

[7] Schwierz F and Liou J J 2007 RF transistors: Recent developments and roadmap toward terahertz applications *Solid. State. Electron.* **51** 1079–91

[8] Ellinger F 2008 *Radio Frequency Integrated Circuits and Technologies* (Springer Publishing Company, Incorporated)

[9] Passi V and Raskin J-P 2017 Review on analog/radio frequency performance of advanced silicon MOSFETs *Semicond. Sci. Technol.* **32** 123004

[10] Lemme M C, Echtermeyer T J, Baus M and Kurz H 2007 A Graphene Field-Effect Device *IEEE Electron Device Lett.* **28** 282–4

[11] Schwierz F 2013 Graphene Transistors: Status, Prospects, and Problems *Proc. IEEE* **101** 1567–84

[12] Ferrari A C, *et al.* 2015 Science and technology roadmap for graphene, related two-dimensional crystals, and hybrid systems *Nanoscale* **7** 4598–810

[13] Cheng R, Bai J, Liao L, Zhou H, Chen Y, Liu L, Lin Y-C, Jiang S, Huang Y and Duan X 2012 High-frequency self-aligned graphene transistors with transferred gate stacks *Proc. Natl. Acad. Sci. U. S. A.* **109** 11588–92

[14] Wu Y, Zou X, Sun M, Cao Z, Wang X, Huo S, Zhou J, Yang Y, Yu X, Kong Y, Yu G, Liao L and Chen T 2016 200 GHz Maximum Oscillation Frequency in CVD Graphene Radio Frequency Transistors *ACS Appl. Mater. Interfaces* **8** 25645–9

[15] Bonmann M, Asad M, Yang X, Generalov A, Vorobiev A, Banszerus L, Stampfer C, Otto M, Neumaier D and Stake J 2019 Graphene Field-Effect Transistors With High Extrinsic $f_T$ and $f_{max}$ *IEEE Electron Device Lett.* **40** 131–4

[16] Yu C, He Z Z, Liu Q B, Song X B, Xu P, Han T T, Li J, Feng Z H and Cai S J 2016 Graphene Amplifier MMIC on SiC Substrate *IEEE Electron Device Lett.* **37** 684–7

[17] Andersson M A, Zhang Y and Stake J 2017 A 185-215-GHz Subharmonic Resistive Graphene FET Integrated Mixer on Silicon *IEEE Trans. Microw. Theory Tech.* **65** 165–72

[18] Zhang Y, Andersson M A and Stake J 2016 A 200 GHz CVD graphene FET based resistive subharmonic mixer *2016 IEEE MTT-S International Microwave Symposium (IMS)* (IEEE) pp 1–4

[19] Habibpour O, He Z S, Strupinski W, Rorsman N, Ciuk T, Ciepielewski P and Zirath H 2017 A W-band MMIC Resistive Mixer Based on Epitaxial Graphene FET *IEEE Microw. Wirel. Components Lett.* **27** 168–70

[20] Habibpour O, Vukusic J and Stake J 2013 A 30-GHz integrated subharmonic mixer based on a multichannel graphene FET *IEEE Trans. Microw. Theory Tech.* **61** 841–7

[21] Yeh C H, Teng P Y, Chiu Y C, Hsiao W T, Hsu S S H and Chiu P W 2019 Gigahertz Field-Effect Transistors with CMOS-Compatible Transfer-Free Graphene *ACS Appl. Mater. Interfaces* **11** 6336–43

[22] Hongming L V, Wu H, Liu J, Niu J, Jiahan Y, Huang C, Li J, Xu Q, Wu X and Qian H 2014 Monolithic graphene frequency multiplier working at 10GHz range *Proceedings of Technical Program - 2014 International Symposium on VLSI Technology, Systems and Application (VLSI-TSA)* vol 332 (IEEE) pp 1–2

[23] Habibpour O, He Z S, Strupinski W, Rorsman N, Ciuk T, Ciepielewski P and Zirath H 2016 Graphene FET Gigabit ON –

OFF Keying Demodulator at 96 GHz *IEEE Electron Device Lett.* **37** 333–6

[24] Harada N, Yagi K, Sato S and Yokoyama N 2010 A polarity-controllable graphene inverter *Appl. Phys. Lett.* **96** 012102

[25] Kim W, Riikonen J, Li C, Chen Y and Lipsanen H 2013 Highly tunable local gate controlled complementary graphene device performing as inverter and voltage controlled resistor *Nanotechnology* **24** 395202

[26] Chen H-Y and Appenzeller J 2012 Graphene-based frequency tripler. *Nano Lett.* **12** 2067–70

[27] Cheng C, Huang B, Mao X, Zhang Z, Zhang Z, Geng Z, Xue P and Chen H 2017 A graphene based frequency quadrupler *Sci. Rep.* **7** 46605

[28] Habibpour O 2012 *Graphene FETs in Microwave Applications* (Chalmers University of Technology)

[29] Zimmer T and Frégonèse S 2015 Graphene Transistor-Based Active Balun Architectures *IEEE Trans. Electron Devices* **62** 3079–83

[30] Li S L, Miyazaki H, Kumatani A, Kanda A and Tsukagoshi K 2010 Low operating bias and matched input-output characteristics in graphene logic inverters *Nano Lett.* **10** 2357–62

[31] Kim W, Li C, Chekurov N, Arpiainen S, Akinwande D, Lipsanen H and Riikonen J 2015 All-Graphene Three-Terminal-Junction Field-Effect Devices as Rectifiers and Inverters *ACS Nano* **9** 5666–74

[32] Champlain J G 2011 A first principles theoretical examination of graphene-based field effect transistors *J. Appl. Phys.* **109** 084515

[33] Feijoo P C, Jiménez D and Cartoixà X 2016 Short channel effects in graphene-based field effect transistors targeting radio-frequency applications *2D Mater.* **3** 025036

[34] Wu Y, Farmer D B, Zhu W, Han S-J, Dimitrakopoulos C D, Bol A A, Avouris P and Lin Y-M 2012 Three-terminal graphene negative differential resistance devices *ACS Nano* **6** 2610–6

[35] Radosavljević M, Heinze S, Tersoff J and Avouris P 2003 Drain voltage scaling in carbon nanotube transistors *Appl. Phys. Lett.* **83** 2435–7

[36] Guo J and Alam M A 2005 Carrier transport and light-spot movement in carbon-nanotube infrared emitters *Appl. Phys. Lett.* **86** 1–4

[37] Yu-Ming Lin, Hsin-Ying Chiu, Jenkins K a., Farmer D B, Avouris P and Valdes-Garcia A 2010 Dual-Gate Graphene FETs With $f_T$ of 50 GHz *IEEE Electron Device Lett.* **31** 68–70

[38] Meric I, Han M Y, Young A F, Oezyilmaz B, Kim P, Shepard K L, Ozyilmaz B, Kim P and Shepard K L 2008 Current saturation in zero-bandgap, top-gated graphene field-effect transistors *Nat. Nanotechnol.* **3** 654–9

[39] Yeh C-H, Lain Y-W, Chiu Y-C, Liao C-H, Moyano D R, Hsu S S H and Chiu P-W 2014 Gigahertz Flexible Graphene Transistors for Microwave Integrated Circuits *ACS Nano* **8** 7663–70

[40] Han S J, Chen Z, Bol A a. and Sun Y 2011 Channel-length-dependent transport behaviors of graphene field-effect transistors *IEEE Electron Device Lett.* **32** 812–4

[41] Han S J, Valdes-Garcia A, Bol A A, Franklin A D, Farmer D, Kratschmer E, Jenkins K A and Haensch W 2011 Graphene technology with inverted-T gate and RF passives on 200 mm platform *Tech. Dig. - Int. Electron Devices Meet. IEDM* 19–22

[42] Wang S, Jin Z, Huang X, Peng S, Zhang D and Shi J 2016 Abnormal Dirac point shift in graphene field-effect transistors *Mater. Res. Express* **3** 095602

[43] Saeed M, Hamed A, Fan C-Y, Heidebrecht E, Negra R, Shaygan M, Wang Z and Neumaier D 2017 Millimeter-wave graphene-based varactor for flexible electronics *2017 12th European Microwave Integrated Circuits Conference (EuMIC)* (IEEE) pp 117–20

[44] Moon J S, Antcliffe M, Seo H C, Curtis D, Lin S, Schmitz A, Milosavljevic I, Kiselev A A, Ross R S, Gaskill D K, Campbell P M, Fitch R C, Lee K-M and Asbeck P 2012 Ultra-low resistance ohmic contacts in graphene field effect transistors *Appl. Phys. Lett.* **100** 203512

[45] Cusati T, Fiori G, Gahoi A, Passi V, Lemme M C, Fortunelli A and Iannaccone G 2017 Electrical properties of graphene-metal contacts *Sci. Rep.* **7** 5109

[46] Chaves F A, Jiménez D, Sagade A A, Kim W, Riikonen J, Lipsanen H and Neumaier D 2015 A physics-based model of gate-tunable metal–graphene contact resistance benchmarked against experimental data *2D Mater.* **2** 025006

[47] Driussi F, Venica S, Gahoi A, Kataria S, Lemme M C and Palestri P 2020 Dependability Assessment of Transfer Length Method to Extract the Metal-Graphene Contact Resistance *IEEE Trans. Semicond. Manuf.* **33** 210–5

[48] Gahoi A, Kataria S, Driussi F, Venica S, Pandey H, Esseni D, Selmi L and Lemme M C 2020 Dependable Contact Related Parameter Extraction in Graphene–Metal Junctions *Adv. Electron. Mater.* **2000386** 1–9

[49] Wang Z, Uzlu B, Shaygan M, Otto M, Ribeiro M, Marin E G, Iannaccone G, Fiori G, Elsayed M S, Negra R and Neumaier D 2019 Flexible One-Dimensional Metal-Insulator-Graphene Diode *ACS Appl. Electron. Mater.* **1** 945–50

[50] Toral-Lopez A, Marin E G, Pasadas F, Gonzalez-Medina J M, Ruiz F G, Jiménez D and Godoy A 2019 GFET Asymmetric Transfer Response Analysis through Access Region Resistances *Nanomaterials* **9** 1027

[51] Urban F, Lupina G, Grillo A, Martucciello N and Di Bartolomeo A 2020 Contact resistance and mobility in back-gate graphene transistors *Nano Express* **1** 010001

[52] Giubileo F and Di Bartolomeo A 2017 The role of contact resistance in graphene field-effect devices *Prog. Surf. Sci.* **92** 143–75

[53] Pasadas F and Jiménez D 2016 Large-Signal Model of Graphene Field-Effect Transistors - Part I: Compact Modeling of GFET Intrinsic Capacitances *IEEE Trans. Electron Devices* **63** 2936–41

[54] Pacheco-Sanchez A and Jiménez D 2019 Efficient contact resistance extraction from individual device characteristics of graphene FETs *2019 IEEE 14th Nanotechnol. Mater. Devices Conf. NMDC 2019* 5–8

[55] Pacheco-Sanchez A, Feijoo P C and Jiménez D 2020 Contact resistance extraction of graphene FET technologies based on individual device characterization *Solid. State. Electron.* **172** 107882

[56] Voinigescu S 2013 *High-Frequency Integrated Circuits* (Cambridge University Press)

[57] Fadil D, Passi V, Wei W, Salk S Ben, Zhou D, Strupinski W, Lemme M C, Zimmer T, Pallecchi E, Happy H and Fregonese S 2020 A broadband active microwave monolithically integrated circuit balun in graphene technology *Appl. Sci.* **10** 2183

[58] Berrie J 2005 Symmetry is central to differential pairs https://www.edn.com/symmetry-is-central-to-differential-pairs/

[59] Bucci D 2011 The mathematical symmetry of the differential pair https://www.electroyou.it/darwinne/wiki/the-differential-pair-and-the-symmetry

[60] Pasadas F and Jiménez D 2016 Large-Signal Model of Graphene Field- Effect Transistors—Part II: Circuit Performance Benchmarking *IEEE Trans. Electron Devices* **63** 2942–7

[61] Peng P, Tian Z, Li M, Wang Z, Ren L and Fu Y 2019 Frequency multiplier based on back-gated graphene FETs with M-shaped resistance characteristics *J. Appl. Phys.* **125** 064503